\documentclass[11pt]{article}
\begin{document}
  
\title{{\bf Dispersionless sTB}}
\author{Ashok Das\\
Department of Physics and Astronomy,\\
University of Rochester,\\
Rochester, NY 14627-0171\\
USA\\
\\
and\\
\\
Ziemowit Popowicz \\
Institute of Theoretical Physics, \\
University of Wroc/l aw,\\
50-205 Wroc/l aw\\ 
Poland.}
\date{}
\maketitle

\begin{abstract}

We analyze the dispersionless limits of the 
SUSY TB-B (sTB-B) and the SUSY TB (sTB) hierarchies. We present the Lax
description for each of these models, as well as the $N=2$ sTB
hierarchy and bring out various properties associated with them. We
also discuss open questions that need to be addressed in connection with
these models. 

\end{abstract}

\vfill\eject
\section{Introduction:}

In recent years, dispersionless integrable models have received a lot
of attention. They involve equations of hydrodynamic type [1-5] and
include such systems as the Riemann equation [5-7], the
the polytropic gas dynamics [5,8], the chaplygin gas  and the
Born-Infeld equation [9-10]. These are models which can be obtained
from a \lq\lq classical'' limit [3,11] of integrable models where the
dispersive terms are absent. They have many interesting properties including
the fact that, unlike their dispersive counterparts, each of them can be
described by a Lax equation which involves a Lax function in the
classical phase space and a classical Poisson bracket relation. Even
more interesting and more difficult are the supersymmetric
dispersionless models. In a recent paper [12], we gave, for the first time,
the Lax description for the dispersionless supersymmetric KdV equation [13]
as well as the dispersionless
Kupershmidt equation [14]. Unlike the bosonic models, the Lax functions for
the dispersionless supersymmetric models do not follow trivially
from the Lax operator of the dispersive counterpart. Furthermore,
while a lot of the interesting properties follow from the Lax
description of the model, we also pointed out several open questions
that arise. In this paper, we follow up on our earlier investigation
and describe the Lax formulation for the dispersionless supersymmetric
two boson (TB) hierarchy.

The TB hierarchy [15] as well as its supersymmetric counterpart [16] are known
to yield various other integrable models upon appropriate
reduction. In this sense, the supersymmetric TB hierarchy is a more
interesting model to study. In particular, it has a natural $N=2$
supersymmetry [16-17] and its dispersionless limit would lead to the
first $N=2$ supersymmetric model of its kind. There are, in fact,  two distinct
supersymmetric generalizations of the TB hierarchy. The first is known
as the sTB-B hierarchy [18], so called because it leads upon reduction to the 
supersymmetric KdV equation considered by the Beckers [19]. The second
supersymmetric generalization, on the other hand, leads upon reduction
to the supersymmetric KdV equation considered by Manin and Radul [13]. We
call this the sTB hierarchy [16] (although, a more appropriate name may be
sTB-MR along the same lines).

In section {\bf 2}, we give the Lax
description for the dispersionless sTB-B hierarchy, which is fairly
straightforward, and bring out its properties as well as the open
questions associated with this model. In section {\bf 3}, we give the
Lax description for the dispersionless sTB hierarchy. This is quite
nontrivial  and naturally reduces to the dispersionless sKdV equation
upon appropriate restriction. However, unlike the sTB model, it does
not lead to the dispersionless
supersymmetric non-linear Schr\"{o}dinger equation. In fact, even the
dispersionless bosonic TB model does not quite give the dispersionless
non-linear
Schr\"{o}dinger equation (at least, we have not succeeded in finding a
field redefinition which would do this). We bring out various
properties of the dispersionless sTB model as well as the open
questions associated with this system. In section {\bf 4}, we describe
the dispersionless sTB system in a manifestly $N=2$ supersymmetric
formulation. Finally,
we make some brief observations in section {\bf 5}.  
We have used REDUCE [27] and special supersymmetric 
package in Reduce [28] extensively in some of the algebraic
calculations.

\section{Dispersionless Limit of sTB-B Equation:}

The sTB hierarchy [16], like the TB hierarchy, is an integrable system in
$1+1$ dimensions. The basic dynamical variables for this system are
the two  fermionic superfields
\begin{eqnarray}
\Phi_{0}(t,x,\theta) & = & \psi_{0} + \theta J_{0}\nonumber\\
\Phi_{1}(t,x,\theta) & = & \psi_{1} + \theta J_{1}
\end{eqnarray}
where $\theta$ represents a Grassmann coordinate and we are
suppressing the space-time dependence on the right hand side for
simplicity. The sTB-B hierarchy is described by the non-standard Lax
equation [18]
\begin{equation}
{\partial L\over \partial t_{n}} = \left[L, (L^{n})_{\geq 1}\right]
\end{equation}
where $n=1,2,...$ and the Lax operator has the form
\begin{equation}
L = D^{2} - (D\Phi_{0}) + D^{-2}(D\Phi_{1}),
\end{equation}
with the super-covariant derivative defined to be
\begin{equation}
D = {\partial\over \partial\theta} + \theta {\partial\over \partial
x},\quad\quad D^{2} = {\partial\over \partial x}
\end{equation}

Explicitly, the first three flows of the sTB-B hierarchy have the form
\begin{equation}
{\partial\Phi_{0}\over \partial t}  =  - \Phi_{0x}, \qquad\qquad
{\partial\Phi_{1}\over \partial t}  = - \Phi_{1x} ,
\end{equation}
\begin{eqnarray}
{\partial\Phi_{0}\over \partial t} & = & - (D^{4}\Phi_{0}) +
D\left((D\Phi_{0})^{2} + 2 (D\Phi_{1}))\right),\nonumber\\
{\partial\Phi_{1}\over \partial t} & = &  (D^{4}\Phi_{1}) + 2
D\left((D\Phi_{0})(D\Phi_{1})\right),
\end{eqnarray}
\begin{eqnarray}
{\partial\Phi_0\over \partial t} & = & -\Phi_{0xxx} -D\left(
6(D\Phi_{0})(D\Phi_1) -3(D\Phi_{0})(D\Phi_{0x}) 
 + (D\Phi_0)^{3}\right), \nonumber\\
{\partial\Phi_{1}\over \partial t} & = &  -\Phi_{1xxx} - 3
D\left((D\Phi_{1})^{2} + (D\Phi_{1})(D\Phi_{0})^{2} +
(D\Phi_{1x})(D\Phi_{0}) - 2(D\Phi_{1})(D\Phi_{0x})\right).
\end{eqnarray}
The last two equations lead, under the reduction $\Phi_0 =0$, to the
supersymmetric  KdV equation considered by the Beckers [19] (the so called
sKdV-B equation). 
\begin{eqnarray}
{\partial\Phi_{1}\over \partial t} & = & -\Phi_{1xxx} -
3D\left((D\Phi_1)^{2}\right).
\end{eqnarray}

The Lax description for the dispersionless limit of the sTB-B hierarchy is
quite straightforward, much like the dispersionless limit of the sKdV-B
hierarchy [12]. Consider the Lax function
\begin{equation}
L = p - (D\Phi_{0}) + p^{-1}(D\Phi_{1})
\end{equation}
where $p$ is the momentum variable of the classical phase space,
satisfying  the canonical PB relations
\begin{equation}
\{x,p\} = 1, \quad \{x,x\} = 0 = \{p,p\}.
\end{equation}  
Then, it is easily seen, with the standard canonical Poisson bracket
relations  ($ \{p,f \} = -\frac{d f}{d x} $), that the Lax equation
\begin{equation}
{\partial L\over \partial t_n} = \left\{(L^{n})_{\geq 1}, L\right\}
\end{equation}
leads to the dispersionless sTB-B hierarchy.  Explicitly, the first
three flows of this hierarchy are ,
\begin{equation}
{\partial\Phi_{0}\over \partial t}  =  -\Phi_{0x},\qquad\qquad
{\partial\Phi_{1}\over \partial t}  = - \Phi_{1x}, 
\end{equation}
\begin{equation}
{\partial\Phi_{0}\over \partial t} = 2(D\Phi_{0})\Phi_{0x} +
2\Phi_{1x},\qquad
{\partial\Phi_{1}\over \partial t} = 2 D((D\Phi_{0})(D\Phi_{1})),
\end{equation}
\begin{eqnarray}
{\partial\Phi_{0}\over \partial t} & = & D\left(-(D\Phi_{0})^{3}
-6(D\Phi_{1})(D\Phi_0)\right)\nonumber\\
{\partial\Phi_{1}\over \partial t} & = & -3D\left((D\Phi_{1})^{2} +
(D\Phi_{1})(D\Phi_0)^{2}\right).
\end{eqnarray}
Equation (14) allows the reduction ($ \Phi_0=0$) to the
dispersionless supersymmetric KdV-B equation
\begin{equation}
{\partial\Phi_{1}\over \partial t} = -3D((D\Phi_1)^{2}).
\end{equation}
whose properties we have studied earlier in detail [12].

From the Lax description, the conserved quantities of the hierarchy
can be easily obtained. In fact, the normalized conserved quantities
can be written as
\begin{eqnarray}
H_{n} & = & {(-1)^{n+1}\over n} \int dz\,{\rm Res}\,L^{n}\nonumber\\
 & = & {1\over n} \sum_{m=0}^{n}\, ^{n}C_{n-m}\,^{n-m}C_{m+1} \int
 dz\,(D\Phi_{0})^{n-2m-1} (D\Phi_{1})^{m+1}
\end{eqnarray}
Here $dz=dxd\theta$ represents the integration over the superspace. Explicitly, the
first few conserved quantities are
\begin{eqnarray*}
H_{1} & = & \int dz\,(D\Phi_{1}) = 0\\
H_{2} & = & \int dz\, (D\Phi_{0})(D\Phi_{1})\\
H_{3} & = & \int dz\, [(D\Phi_{0})^{2}+(D\Phi_{1})](D\Phi_{1})
\end{eqnarray*}
and so on. As it stands, it is clear that these conserved quantities are
fermionic. This is a peculiarity of the sTB-B system (for that matter any -B
system) that the Hamiltonians are fermionic. Correspondingly, the
Hamiltonian structures are odd and we note the first two structures for
completeness, namely,
\begin{equation}
{\cal D}_{1} = \left(\begin{array}{cc}
                          0 & 1\\
                          1 & 0
                          \end{array}\right),\quad
			  {\cal D}_{2} = \left(\begin{array}{cc}
                          2 & D(D\Phi_{0})D^{-1}\\
                          D^{-1}(D\Phi_{0})D &
                          D^{-1}(D\Phi_{1})D+D(D\Phi_{1})D^{-1}
                          \end{array}\right)
\end{equation}
While the Hamiltonian structure ${\cal D}_{1}$ can be obtained from
the Lax function as the standard Gelfand-Dikii bracket, we do not know
how to obtain the second structure from the Lax function. Furthermore,
the Jacobi identity for this structure is complicated and needs to be
checked. However, these two Hamiltonian structures do lead to the
recursion operator
\begin{equation}
R = {\cal D}_{1}^{-1}{\cal D}_{2} = \left(\begin{array}{cc}
                                          D^{-1}(D\Phi_{0})D &
                                       D^{-1}(D\Phi_{1})D+D(D\Phi_{1})D^{-1}\\
                                          2 & D(D\Phi_{0})D^{-1}
                                          \end{array}\right)\label{recursion}
\end{equation}
which can be easily checked to connect the successive Hamiltonians of
the hierarchy.

Supersymmetric integrable systems have conserved non-local charges
[20-21] and the dispersionless sTB-B hierarchy also has conserved
non-local charges. For example, it can be checked that
\begin{eqnarray}
Q_{n} & = & {(-1)^{n+1}\over n} \int dz\,(D^{-1}{\rm
Res}\,L^{n})\nonumber\\
 & = & {1\over n} \sum_{m=0}^{n} \,^{n}C_{n-m}\,^{n-m}C_{m+1} \int
dz\,(D^{-1}((D\Phi_{0})^{n-2m-1}(D\Phi_{1})^{m+1}))
\end{eqnarray}
with the first few charges
\begin{eqnarray}
Q_{1} & = & \int dz\,\Phi_{1}\nonumber\\
Q_{2} & = & \int dz\,(D^{-1}((D\Phi_{0})(D\Phi_{1})))\nonumber\\
Q_{3} & = & \int dz\,(D^{-1}((D\Phi_{0})^{2}(D\Phi_{1}) +
(D\Phi_{1})^{2}))
\end{eqnarray}
and so on, are conserved under the flow of the system. These are
bosonic charges and, interestingly enough, these non-local charges are
also related to one another by the same recursion operator $R$ in
eq. (\ref{recursion}), namely,
\begin{equation}
\left(\begin{array}{c}
       {\delta Q_{n+1}\over \delta\Phi_{0}}\cr
       {\delta Q_{n+1}\over \delta\Phi_{1}}
      \end{array}\right) = R\,\left(\begin{array}{c}
                                     {\delta Q_{n}\over
       \delta\Phi_{0}}\cr
       {\delta Q_{n}\over \delta\Phi_{1}}
       \end{array}\right)
\end{equation}
It is also nice to see explicitly that if we set $\Phi_{0}=0$ and
$\Phi_{1}=\Phi$, then, all the even charges vanish and the odd
charges, namely, $H_{2n+1}$ and $Q_{2n+1}$ coincide with the
corresponding charges of the sKdV-B system [12], as they should.

\section{Dispersionless Limit of sTB Equation:}

In terms of the same basic variables, $\Phi_{0}$ and $\Phi_{1}$, the
sTB hierarchy is described by the Lax operator [16]
\begin{equation}
L = D^{2} - (D\Phi_{0}) + D^{-1}\Phi_{1},
\end{equation}
and the non-standard Lax equation
\begin{equation}
{\partial L\over \partial t_n} = \left[L, (L^{n})_{\geq 1}\right].
\end{equation}
Explicitly, the lowest order equations have the form
\begin{equation}
{\partial\Phi_{0}\over \partial t}  =  - \Phi_{0x},\qquad\qquad
{\partial\Phi_{1}\over \partial t}  = - \Phi_{1x}, 
\end{equation}
\begin{eqnarray}
{\partial\Phi_{0}\over \partial t} & = & - (D^{4}\Phi_{0}) +
D((D\Phi_{0})^{2} + 2 (D\Phi_{1})),\nonumber\\
{\partial\Phi_{1}\over \partial t} & = &  (D^{4}\Phi_{1}) + 2D^{2}\left((D\Phi_{0})\Phi_{1}\right),
\end{eqnarray}
\begin{eqnarray}
{\partial\Phi_{0}\over \partial t} & = & -(D^{6}\Phi_{0}) +
D\left(3\Phi_{1}\Phi_{0x}-6(D\Phi_{1})(D\Phi_{0})-(D\Phi_{0})^{3}+3(D\Phi_{0})(D\Phi_{0x})\right),\nonumber\\
{\partial\Phi_{1}\over \partial t} & = & -(D^{6}\Phi_{1}) - 3D^{2}\left(\Phi_{1}(D\Phi_{0})^{2}+(D\Phi_{0})\Phi_{1x}+\Phi_{1}(D\Phi_{1})\right).
\end{eqnarray}
The last equation allows the reduction ($\Phi_0=0$) to the
supersymmetric KdV equation considered by Manin and Radul [13], namely,
\begin{equation}
{\partial\Phi_{1}\over \partial t}  = -(D^{6}\Phi_{1}) -
3D^{2}\left(\Phi_{1}(D\Phi_{1})\right) 
\end{equation}

The dispersionless limit of this Lax operator is not as
straightforward. However, with some work, it can be determined that
the Lax function
\begin{equation}
L = p - (D\Phi_{0}) + p^{-1}(D\Phi_{1}) - p^{-2}\Phi_{0x}\Phi_{1} +
p^{-3}\Phi_{1x}\Phi_{1} 
\end{equation}
and the classical Lax equation
\begin{equation}
{\partial L\over \partial t_{n}} = \left\{(L^{n})_{\geq 1}, L\right\}
\end{equation}
give the dispersionless sTB hierarchy whose first three
equations have the explicit forms
\begin{equation}
{\partial\Phi_{0}\over \partial t}  =  -\Phi_{0x}, 
\qquad\qquad
{\partial\Phi_{1}\over \partial t} =  -\Phi_{1x},
\end{equation}
\begin{equation}
{\partial\Phi_{0}\over \partial t} = 2(D\Phi_{0})\Phi_{0x} +
2\Phi_{1x},\qquad
{\partial\Phi_{1}\over \partial t} = 2D^{2}\left((D\Phi_{0})\Phi_{1}\right)
\end{equation}
\begin{eqnarray}
{\partial\Phi_{0}\over \partial t} & = &
D\left(3\Phi_{1}\Phi_{0x}-6(D\Phi_{1})(D\Phi_0) -
(D\Phi_{0})^{3}\right),\nonumber\\
{\partial\Phi_{1}\over \partial t} & = &
-3D^{2}\left(\Phi_1(D\Phi_0)^2 + \Phi_1(D\Phi_{1})\right).
\end{eqnarray}

There are several things to note from here. First, eq. (32),
upon setting $\Phi_{0}=0$ and $\Phi_{1}=-\Phi$, gives the
dispersionless sKdV equation [12], namely,
\begin{equation}
{\partial\Phi\over \partial t} =  3D^{2}\left(\Phi(D\Phi)\right).
\end{equation}
This, therefore, gives
the non-standard representation of the dispersionless sKdV equation
(as opposed to the standard representation given in [12]),
and is analogous to the reduction of the sTB hierarchy to the sKdV
hierarchy. Second, the normalized conserved quantities of this system can be
easily determined to be
\begin{eqnarray}
H_{n} & = & {(-1)^{n+1}\over n}\,\int dz\,D^{-1}({\rm
Res}L^{n})\nonumber\\
 & = & \sum_{m=0}^{m_{max}} {(n-m)!\over m!(m-1)!(n-2m+1)!}\,\int
dz\,(D\Phi_{0})^{n-2m+1}(D\Phi_{1})^{m-1}\Phi_{1}
\end{eqnarray}
where the upper limit $m_{max}={n\over 2}$ if $n$ is even, but
$m_{max}=[{n\over 2}]+1$ if $n$ is odd. The first few of these
conserved charges have the explicit forms
\begin{eqnarray}
H_{1} & = & \int dz\,\Phi_{1}\nonumber\\
H_{2} & = & \int dz\,(D\Phi_{0})\Phi_{1}\nonumber\\
H_{3} & = & \int dz\, ((D\Phi_{0})^{2} + (D\Phi_{1}))\Phi_{1}
\end{eqnarray}
and so on. These conserved charges are all bosonic and it is clear
that if we set $\Phi_{0}=0$ and $\Phi_{1}=\Phi$, all the even charges
vanish while the odd charges coincide with those of the dispersionless
sKdV hierarchy [12] as they should.

We have not been able to derive the Hamiltonian structures for this
system from the Gelfand-Dikii formalism (as is the case in the
dispersionless sKdV also[12]). However, the first Hamiltonian structure can be
easily checked to be
\begin{equation}
{\cal D}_{1} = \left(\begin{array}{cc}
                   0 & D\cr
                   D & 0
                   \end{array}\right)
\end{equation}
and this trivially satisfies the Jacobi identity. Notice that this
Hamiltonian operator  defines a closed skew symmetric two-form 
\begin{equation}
\Omega( {\cal D}_1) (a,b) = \int dz (a_1(Db_2) + a_2(Db_1))
\end{equation}
where a,b are arbitrary, two component (column matrix) bosonic superfields. 
We have also checked that it is impossible 
to construct the second Hamiltonian operator out of local
operators alone.  However, there is  a possibility, which we have not
checked,  to construct such an operator out of local as well as
non-local operators. 

Let us now discuss some of the outstanding questions associated with
such a system. First, we have not been able to construct the non-local
charges from this Lax function. By brute force, we have checked that
charges, such as
\begin{eqnarray}
Q_{n} & = & \int dz\,(D^{-1}\Phi_{1})^{n},\qquad\qquad
n=1,2,3,\cdots\nonumber\\
Q'_{2} & = & \int dz\,\left[{3\over 2} (D^{-1}\Phi_{1})^{2} + {1\over
2} \Phi_{0}\Phi_{1} - (D^{-1}((D\Phi_{0})\Phi_{1}))\right]
\end{eqnarray}
and so on, are conserved under the flow. Furthermore, under the
substitution, $\Phi_{0}=0$ and $\Phi_{1}=\Phi$, these reduce to the
appropriate non-local charges of the dispersionless sKdV [12] (In this
limit, $Q_{2}={2\over 3}Q'_{2}$). However, for
lack of a systematic procedure for constructing these charges, we do
not have a general expression for the $n$th charge of this
set in terms of the Lax function. Furthermore, we do know that the
dispersionless sKdV has a second set of non-local charges [12] and since
this system can be obtained from the 
dispersionless sTB, it would be natural to expect the dispersionless
sTB also to have a second set of non-local charges. However, we do not
know if such a set exists. This emphasizes the need for a systematic
understanding of the non-local charges in such systems, described a
classical Lax function.

It is also worth recalling that the sTB equation yields the sNLS equation
(supersymmetric non-linear Schr\"{o}dinger equation) under the
redefinition [16]
\begin{equation}
\Phi_{0} = - (D\ln(DQ))+(D^{-1}(\bar{Q}Q)),\quad \Phi_{1} = -
\bar{Q}(DQ)
\end{equation}
However, we would like to note here that we have not been able to find
a redefinition of fields which would take the dispersionless sTB
equation to the dispersionless sNLS equation. In fact, it is worth
pointing out that it is not a difficulty only for the supersymmetric
system. Even the dispersionless TB equation (bosonic) does not appear
to yield the dispersionless NLS equation under the corresponding
redefinition. This question certainly needs further study.

\section{$N=2$ Formulation for the dispersionless sTB Equation:}

The sTB hierarchy has a $N=2$
supersymmetry [16]. However, it is not manifest in the description of the
last section. In order to see the $N=2$ supersymmetry manifestly,
we have to define the basic variables of the theory appropriately. Let
us recall  that in the conventional description of the system (as
given in the previous section), the basic variables were two fermionic
superfields which depended on the usual bosonic coordinates, $(x,t)$,
but they also depended on an additional anti-commuting  Grassmann
variable $\theta$. The Taylor expansion of such superfields in the
Grassmann coordinate  is simple
and has been given in eq. (1). In the presence of an $N=2$
supersymmetry, the superfields depend on  $(x,t)$ as well as two
anti-commuting  variables $\theta_1$ and
$\theta_2$. (Incidentally, an $N=2$ supersymmetric system can also be
described on an $N=1$ superspace which is the description used in the
earlier section.) Expanding the superfield
$\Phi(x,t,\theta_1,\theta_2)$ in  a Taylor series in the
anti-commuting variables, we obtain
\begin{equation}
\Phi =\phi_1 + \theta_1\chi_1 +\theta_2\chi_2 + \theta_2\theta_1\phi_2,
\end{equation}
where $\phi_1,\phi_2$ are bosonic functions (or fermionic ) and
$\chi_1,\chi_2$  are fermionic (or bosonic ) if $\Phi$ is a
bosonic (or fermionic) superfield. In this case, the super-covariant
derivatives on the $N=2$ superspace are defined as
\begin{eqnarray}
D_1 & = & \frac{\partial}{\partial \theta_1} + \theta_1{\partial\over
\partial x}, {~~~} 
D_2  =  \frac{\partial}{\partial \theta_2} + \theta_2{\partial\over
\partial x},\nonumber\\
D_1^{2}& =& D_2^{2}= \partial = {\partial\over \partial x} ,{~~~}
D_2D_1 = - D_1D_2.
\end{eqnarray}

Recently several methods have been proposed to obtain
$N=2$ supersymmetric soliton equations [22-25]. In this paper, we consider two
different supersymmetric  Lax operators which generate two distinct
sTB hierarchies. 
The first is connected with the $N=2, a=1$ [22,24] supersymmetric KdV equation
(namely, it reduces to this model upon appropriate redefinition) and
is defined by the Lax operator
\begin{equation}
L=\partial + \tilde{\Phi}_0 + \partial^{-1}D_1D_2\tilde{\Phi}_1.
\end{equation}
where $\tilde{\Phi}_0$ and $\tilde{\Phi}_1$ are two bosonic
superfields (and we have used a tilde to avoid confusion with the
superfields of the earlier sections). This Lax operator is related
to the supersymmetric Lax operator considered in [24], through the gauge
transformation  
\begin{equation}
L=e^{-g} L e^{g},
\end{equation}
where $g=\int dxd\theta_1d\theta_2\,G=\int dZ\,G$ with $G$  a bosonic
superfield.

It is straightforward to check that the  Lax operator of eq. (42)
leads to consistent dynamical equations through the nonstandard Lax
relation
\begin{equation}
{\partial L\over \partial t_{n}} = \left[L, (L^{n})_{\geq 1}\right].
\end{equation}
Explicitly, we can write the  first three flows as
\begin{equation}
{\partial \tilde{\Phi}_{0} \over \partial t}  =  -\tilde{\Phi}_{0x},
\qquad\qquad {\partial \tilde{\Phi}_1\over \partial t}  =  -\tilde{\Phi}_{1x},
\end{equation}
\begin{equation}
{\partial \tilde{\Phi}_0 \over \partial t} =
\partial(\tilde{\Phi}_{0x}-\tilde{\Phi}_0^2-2(D_1D_2\tilde{\Phi}_1)
+\tilde{\Phi}_1^2),\qquad
{\partial \tilde{\Phi}_1\over \partial t} =
\partial(\tilde{\Phi}_{1x}-2\tilde{\Phi}_1\tilde{\Phi}_0)
\end{equation}
\begin{eqnarray}
{\partial \tilde{\Phi}_0 \over \partial t} & = &
 \partial(-\tilde{\Phi}_{0xx}-3\tilde{\Phi}_{ox}\tilde{\Phi}_0
 -\tilde{\Phi}_0^3 -3(D_1D_2\tilde{\Phi}_1)\tilde{\Phi}_0
 +3\tilde{\Phi}_1^2\tilde{\Phi}_0-3(D_1D_2(\tilde{\Phi}_1\tilde{\Phi}_0))),
 \nonumber\\
{\partial \tilde{\Phi}_1\over \partial t} & = &
 \partial(-\tilde{\Phi}_{1xx} -3(D_1D_2\tilde{\Phi}_1)\tilde{\Phi}_1
 +\tilde{\Phi}_1^3 + 3\tilde{\Phi}_{1x}\tilde{\Phi}_0
 -3\tilde{\Phi}_1\tilde{\Phi}_0^2).
\end{eqnarray}
The last equation leads, upon the reduction $\tilde{\Phi}_0=0$, to the
$N=2,a=1$ [22] supersymmetric KdV equation
\begin{eqnarray}
{\partial \tilde{\Phi}_1\over \partial t}  =
\partial(-\tilde{\Phi}_{1xx} -3(D_1D_2\tilde{\Phi}_1)\tilde{\Phi}_1
+\tilde{\Phi}_1^3 ).
\end{eqnarray}

We can also construct a  second Lax operator of the  form
\begin{equation}
L=D_1 + \overline{\Phi}_0 + \partial^{-1}D_2\overline{\Phi}_1,
\end{equation}  
where $\overline{\Phi}_0$ is a fermionic superfield while
$\overline{\Phi}_1$  is a bosonic superfield. Note that the Lax
operator, in the present case, is fermionic while that in eq. (42) was
bosonic. Nonetheless, as in the previous case, it is easy to check
that this Lax operator is gauge equivalent to the one considered in
[25]. This Lax operator leads to dynamical equations of the non-standard
form  
\begin{equation}
{\partial L\over \partial t_{2n}} = \left[L, (L^{2n})_{\geq 1}\right].
\end{equation}
Explicitly, the first three flows of this hierarchy have the forms 
\begin{equation}
{\partial \overline{\Phi}_0 \over \partial t}  =
-\overline{\Phi}_{0x},\qquad\qquad {\partial \overline{\Phi}_1\over
\partial t}  =  -\overline{\Phi}_{1x},
\end{equation}
\begin{eqnarray}
{\partial \overline{\Phi}_0 \over \partial t} & = & D_1 \big
(-(D_1\overline{\Phi}_{0x})-2(D_1D_2\overline{\Phi}_1)-2(D_1\overline{\Phi}_0)^2
- 2\overline{\Phi}_1^2-\overline{\Phi}_1(D_2\overline{\Phi}_0)\big),\nonumber\\
{\partial \overline{\Phi}_1\over \partial t} & = & D_1 \big
((D_1\overline{\Phi}_{1x})-2\overline{\Phi}_{0x}\overline{\Phi}_1-(D_2\overline
{\Phi}_1)\overline{\Phi}_1
-2(D_1\overline{\Phi}_1)(D_1\overline{\Phi}_0)\big ),
\end{eqnarray}
\begin{eqnarray}
{\partial \overline{\Phi}_0 \over \partial t} & = & D_1 \big
(-(D_1\overline{\Phi}_{0xx}-3(D_1\overline{\Phi}_{0x})(D_1\overline{\Phi}_0)-
6(D_1D_2\overline{\Phi}_1)(D_1\overline{\Phi}_0)-6\overline{\Phi}_1^2(D_1\overline{\Phi}_0)\nonumber\\ 
&& {~~~~~~} -6\overline{\Phi}_1(D_2\overline{\Phi}_0)(D_1\overline{\Phi}_0) +
3(D_2\overline{\Phi}_1\overline{\Phi}_{0x} + 
 3(D_1\overline{\Phi}_1)(D_1D_2\overline{\Phi}_0)
-(D_1\overline{\Phi}_0)^3\big),\nonumber\\
{\partial \overline{\Phi}_1\over \partial t} & = & D_1 \big
(-(D_1\overline{\Phi}_{1xx}+3\overline{\Phi}_{0x}\overline{\Phi}_{1x} 
-6\overline{\Phi}_{0x}\overline{\Phi}_1(D_1\overline{\Phi}_0)+3(D_2\overline{\Phi}_1)\overline{\Phi}_{1x}\nonumber\\
&& {~~~~~~}-
6(D_2\overline{\Phi}_1)\overline{\Phi}_1(D_1\overline{\Phi}_0)+3(D_1\overline{\Phi}_{1x})
(D_1\overline{\Phi}_0)-3(D_1\overline{\Phi}_1)(D_1\overline{\Phi}_0)^2\nonumber\\ 
&& {~~~~~~} -3(D_1\overline{\Phi}_1)(D_1D_2\overline{\Phi}_1)- 
 6(D_1\overline{\Phi}_1)\overline{\Phi}_1^2-6(D_1\overline{\Phi}_1)\overline{\Phi}_1(D_2\overline{\Phi}_0)\big )
\end{eqnarray}
Note that when  $\overline{\Phi}_0 =0$,  the last equation reduces to
the $N=2,a=-2$ [22] supersymmetric KdV equation.
Unfortunately the Lax operator, which gives rise to the sTB hierarchy
that contains  the $N=2, a=4$ [22] supersymmetric KdV equation is not known
as yet. Furthermore, since fermionic Lax operators do not lend easily
to a dispersionless limit, we will not discuss this system any further.

In order to obtain the dispersionless sTB hierarchy, we now have to introduce 
the concept of the fermionic momenta on the $N=2$ superspace. There
will be two such fermionic momenta defined by [12,26]
\begin{equation} 
\Pi_1 =-(p_{\theta_1} + \theta_1p),\qquad\qquad \Pi_2=-(p_{\theta_2} +
\theta_2p),
\end{equation}
We can now assume the 
\lq\lq commutation" rules for the functions $\Pi_{i}$ as 
\begin{equation}
\{\Pi_{i},\Pi_{j}\} = - 2p\delta_{ij}
\end{equation}
Note that the $\Pi_{i}$'s  generate covariant
differentiation through the PB relation (for any superfield $A$)
\begin{equation}
\{\Pi_{i},A\} = (D_{i}A),\qquad\qquad i=1,2. 
\end{equation}

In trying to obtain the dispersionless  $N=2$ sTB hierarchy, we start
from the Lax operator for the $N=2$ sTB hierarchy in eq. (42), and
assume the Lax function for the dispersionless system to have the
general form
\begin{equation}
L=p + k_1\Pi_1\Pi_2 + \tilde{\Phi}_0 + \sum_{s=1}^{3} p^{-s} 
\big ( F_0^{s} + \Pi_1F_1^{s} + \Pi_2F_2^{s} + \Pi_1\Pi_2F_3^{s}\big ).
\end{equation}
where $k_1$ is an arbitrary coefficient and $F_k^{s}
,k=0,1,2,3,s=1,2,3$ are  arbitrary functions of  $\tilde{\Phi}_0$ and
$\tilde{\Phi}_1$. 

We have checked that the classical analogue of eq. (44), namely,
\begin{equation}
{\partial L\over \partial t_{n}} = \left\{(L^{n})_{\geq 1},L\right\}
\end{equation}
with the projection $ \geq 1 $ defined as 
\begin{eqnarray}
&&\Big (\sum_{s=-\infty}^{\infty} p^{s} 
\big ( F_0^{s} + \Pi_1F_1^{s} + \Pi_2F_2^{s} + \Pi_1\Pi_2F_3^{3}\big )\Big)_{\geq 1}=\nonumber\\
&&\Pi_1F_1^{s} + \Pi_2F_2^{s} + \Pi_1\Pi_2F_3^{s}\ +
\sum_{s=1}^{\infty} p^{s} 
\big ( F_0^{s} + \Pi_1F_1^{s} + \Pi_2F_2^{s} + \Pi_1\Pi_2F_3^{3}\big ).
\end{eqnarray}
leads to two possible solutions with $k_1=0$ in either case.

The first Lax function that leads to consistent equations has the form 
\begin{equation}
L=p+\tilde{\Phi}_0+p^{-1}\Pi_1\Pi_2\tilde{\Phi}_1
\end{equation} 
The first three flows of this hierarchy have the explicit forms 
\begin{eqnarray}
{\partial \tilde{\Phi}_0 \over \partial t} & = &  -\tilde{\Phi}_{0x},
\qquad\qquad{\partial  \tilde{\Phi}_1\over \partial t}  =  -\tilde{\Phi}_{1x},
\nonumber\\
{\partial \tilde{\Phi}_0 \over \partial t} & = & \partial
(\tilde{\Phi}_{0}^{2}),\qquad\qquad
{\partial \tilde{\Phi}_1\over \partial t} = 2\partial
(\tilde{\Phi}_1\tilde{\Phi}_0)\nonumber\\
{\partial \tilde{\Phi}_0 \over \partial t} & = & \partial
(\tilde{\Phi}_{0}^{3}),\qquad\qquad
{\partial \tilde{\Phi}_1\over \partial t} = 3\partial
(\tilde{\Phi}_1\tilde{\Phi}_0^2)
\end{eqnarray}
However, this hierarchy appears trivial since the equations do not have
any explicit dependence on supersymmetric covariant
derivatives. Therefore, we will not consider this hierarchy any further. 

The second Lax function does not contain the fermionic functions
$\Pi_i$ and has the form
\begin{equation}
L=p+\tilde{\Phi}_0 + p^{-1}(D_1D_2\tilde{\Phi}_1) -
p^{-2}(D_2\tilde{\Phi}_1)(D_1\tilde{\Phi}_0) +
p^{-3}(D_2\tilde{\Phi}_{1x})(D_2\tilde{\Phi}_1) 
\end{equation}
This, on the other hand, produces an interesting supersymmetric
hierarchy whose first three flows have the forms
\begin{equation}
{\partial \tilde{\Phi}_0 \over \partial t}  =  -\tilde{\Phi}_{0x},
\qquad\qquad{\partial \tilde{\Phi}_1\over \partial t} =  -\tilde{\Phi}_{1x},
\end{equation}
\begin{equation}
{\partial \tilde{\Phi}_0 \over \partial t} = \partial
(2(D_1D_2\tilde{\Phi}_{1})+\tilde{\Phi}_0^{2}),\qquad
{\partial \tilde{\Phi}_1\over \partial t} =
2D_2((D_2\tilde{\Phi}_1)\tilde{\Phi}_0)
\end{equation}
\begin{eqnarray}
{\partial \tilde{\Phi}_0 \over \partial t} & = & \partial
(\tilde{\Phi}_{0}^{3} +6(D_1D_2\tilde{\Phi}_1)\tilde{\Phi}_0 -
3(D_2\tilde{\Phi}_1)(D_1\tilde{\Phi}_0)),\cr
{\partial \tilde{\Phi}_1\over \partial t} & = &
3D_2((D_2\tilde{\Phi}_1)\tilde{\Phi}_0^2 +
(D_2\tilde{\Phi}_1)(D_1D_2\tilde{\Phi}_1)).
\end{eqnarray}
It is interesting to note that, when $\tilde{\Phi}_0=0$, the last
equation  becomes
\begin{equation}
{\partial \tilde{\Phi}_1\over \partial t}  =
3D_2((D_2\tilde{\Phi}_1)(D_1D_2\tilde{\Phi}_1)).
\end{equation}
which, in fact, reduces to eq. (33) with the substitution
$\left.(D_2\tilde{\Phi}_1)\right|_{\theta_{2}=0} = -\Phi_1$. Therefore,
eq. (66) can be  considered as the $N=2$ generalization of the
dispersionless $N=1$ supersymmetric KdV-MR equation. 

Let us also note that the Lax function in eq. (62) as well as the
resulting hierarchy coincide with  those of the previous section with
the identifications
\begin{equation}
\left.(D_2\tilde{\Phi}_1)\right|_{\theta_{2}=0}=-\Phi_1
,{~~~~~~~~~~~~~} \left.\tilde{\Phi}_0\right|_{\theta_{2}=0}
=(D_1\Phi_0) = - (D\Phi_{0}). 
\end{equation}
Namely, the redefinition also involves $\theta_{1}\rightarrow -\theta$ or
equivalently, $D_{1}\rightarrow -D$. Notice that these transformations
are highly nontrivial reductions. Consequently, we do not expect the
conserved quantities of eqs. (34) to define conserved
quantities of the $N=2$ supersymmetric hierarchy and we have verified
this. On the other hand, we can construct conserved quantities for the
$N=2$ hierarchy directly and we have found two such series of conserved
quantities, by brute force, using the computer.

The first set consists of bosonic conserved charges of the form
\begin{eqnarray}
H_n &=& \int dZ\, (\tilde{\Phi}_1)^{n} , {~~~~~~~~~~~~~~~~~~~~~~~~}
n=1,2,3,4,...,\nonumber\\
\tilde{H}_{1} & = & \int dZ\,\tilde{\Phi}_{0}\nonumber\\
\tilde{H}_{2} & = & \int dZ\,\tilde{\Phi}_{1}\tilde{\Phi}_{0}
\end{eqnarray}
where $dZ=dxd\theta_1d\theta_2$. The second series of conserved
charges is fermionic of the form 
\begin{eqnarray}
H_{5/2} &=& \int dZ\, (D_2\tilde{\Phi}_1)\tilde{\Phi}_0\nonumber\\
H_{7/2} &=& \int dZ\, (D_2\tilde{\Phi}_1)(\tilde{\Phi}_0^2 + (D_1D_2\tilde{\Phi}_1)), \nonumber\\
H_{9/2} &=& \int dZ\,
(D_2\tilde{\Phi}_1)(3(D_1D_2\tilde{\Phi}_1)\tilde{\Phi}_0+
\tilde{\Phi}_0^3) , \nonumber\\
H_{11/2} &=& \int dZ\, \big
(4(D_2\tilde{\Phi}_{1x})\tilde{\Phi}_{1x}\tilde{\Phi}_1
-4(D_2\tilde{\Phi}_1)\tilde{\Phi}_{1xx}\tilde{\Phi}_1 + 
12(D_1\tilde{\Phi}_{1x})(D_1D_2\tilde{\Phi}_1)\tilde{\Phi}_1\cr
&& {~~~~~~}+
18(D_2\tilde{\Phi}_1)(D_1D_2\tilde{\Phi}_1)\tilde{\Phi}_0^2 +
3(D_2\tilde{\Phi}_1)\tilde{\Phi}_0^4\big) 
\end{eqnarray}
Here, we have labelled the conserved quantities by the weights of the
integrand ($[\tilde{\Phi}_{0}]=1=[\tilde{\Phi}_{1}]$ and
$[D_{1}]={1\over 2}=[D_{2}]$). Note that the second set of charges in
eq. (69) follows from the Lax function, up to non-essential normalization, as
\begin{equation}
H_{n+{1\over 2}} = \int dZ\,D_{1}^{-1}\,{\rm Res} L^{n}
\end{equation}
However, we do not know how to obtain the first set (except for the
lowest one) from the Lax function, nor is it clear that these exhaust
all the conserved charges of the system.

Comparing with the discussion of the previous section and
particularly from the form of the non-local conserved charges in
eq. (38), we see that we can write non-local conserved charges for
the $N=2$ system as 
\begin{equation} 
\tilde{Q}_n = \int dZ (D_1^{-1}D_2\tilde{\Phi}_1)^{n} \qquad\qquad
n=1,2,3,4\cdots
\end{equation}
It is, in fact, quite straightforward to check that they are conserved.
Similarly, we note that
\begin{equation}
\tilde{Q}'_2=\int dZ (D_1^{-1}(\tilde{\Phi}_0(D_2\tilde{\Phi}_1))
-\frac{1}{2}(D_1^{-1}\tilde{\Phi}_0)(D_2\tilde{\Phi}_1)  -
\frac{3}{2}\tilde{\Phi}_1^{2}),
\end{equation}
also represents a conserved charge (compare with the second of the
charges in eq. (38)). However, we do not know how to obtain these from
the Lax function directly.

The first Hamiltonian structure for the dispersionless $N=2$ hierarchy
of eqs. (63)-(69) is easily seen to be
\begin{equation}
\tilde{{\cal D}}_{1} = \left(\begin{array}{cc}
                   0 & D_2\cr
                   -D_2 & 0
                   \end{array}\right)
\end{equation}
and is trivially seen to satisfy the Jacobi identity. This Hamiltonian
operator defines the closed  skew symmetric two-form 
\begin{equation}
\Omega(\tilde{{\cal D}}_1) (a,b) = \int dZ (a_1(Db_2) - a_2(Db_1))
\end{equation}
where in contrast to the $N=1$ case $a$ and $b$ are arbitrary, two
component fermionic  superfields.

\section{Conclusion}

In this paper, we have studied the dispersionless limits of the sTB-B,
the sTB as well as the $N=2$ sTB hierarchies in detail. We have
obtained the Lax descriptions in terms of classical Lax functions,
obtained conserved local as well as some of the non-local charges and
brought out various other features associated with such systems. We
have also tried to point out various open questions associated with
such systems, the most pressing of which is a systematic understanding
of the construction of non-local charges for such systems starting
from the Lax description as well as a generalization of the
Gelfand-Dikii procedure for construction of Hamiltonian structures for
such systems..

\section*{Acknowledgments}

A.D. acknowledges support in part by the U.S. Dept. of
Energy Grant  DE-FG 02-91ER40685 while Z.P. is supported in part by
the  Polish KBN  Grant  2 P0 3B 136 16.

\end{document}